%% file: main.tex
\documentclass[12pt,a4]{article}
\usepackage[T1]{fontenc}
\usepackage{amsfonts}
\usepackage{amsmath}
\usepackage{amssymb}
\usepackage{amsbsy}
\usepackage{color}
\usepackage{graphicx}
\usepackage{epstopdf}\usepackage{epsfig} 
\usepackage{array}
\usepackage{enumerate}
\usepackage{enumitem}
\usepackage{sectsty}
\usepackage{appendix}
\usepackage{multirow}
\usepackage{cite}
\usepackage{tensor}
\usepackage{algorithm}
\usepackage{algorithmicx}
\usepackage{algpseudocode}
\usepackage{bbold} 
\usepackage{pgf}
\usepackage{tikz}
\usetikzlibrary{shapes,snakes,arrows}
\usepackage{subfigure}
\usepackage{url}
\urldef{\mailsb}\path|e.di.napoli@fz-juelich.de|
\urldef{\mailsa}\path|mberljaf@student.math.hr|
\urldef{\mailsc}\path|dinapoli@aices.rwth-aachen.de|  

\input{symacros}

\begin{document}
%

\title{A Parallel and Scalable Iterative Solver for Sequences of Dense Eigenproblems Arising in FLAPW}

            
\author{Mario Berljafa \footnote{Department of Mathematics, Faculty of
    Science, University of Zagreb.  Bijeni\v{c}ka cesta 30,
    10000--Zagreb, Republic of Croatia.  \mailsa} 
\and Edoardo Di Napoli
\footnote{J\"ulich Supercomputing Centre,
Forschungszentrum J\"ulich.
Wilhelm-Johnen stra\ss e, 52425--J\"ulich, Germany.
\mailsb} \footnote{
Aachen Institute for Advance Study in Computational Engineering Science.
Schinkelstra\ss e 2, 52072 Aachen, Germany. \mailsc}
}%



\maketitle

\begin{abstract} 
  In one of the most important methods in Density Functional Theory --
  the Full-Potential Linearized Augmented Plane Wave (FLAPW) method --
  dense generalized eigenproblems are organized in long
  sequences. Moreover each eigenproblem is strongly correlated to the
  next one in the sequence. We propose a novel approach which exploits
  such correlation through the use of an eigensolver based on subspace
  iteration and accelerated with Chebyshev polynomials. The resulting
  solver, parallelized using the Elemental library framework, achieves
  excellent scalability and is competitive with current dense parallel
  eigensolvers.

\end{abstract}

\pagenumbering{arabic}


\input{body}

\section*{Acknowledgements}

This research was in part supported by the VolkswagenStiftung through
the fellowship "Computational Sciences". We are thankful to the
J\"ulich Supercomputing Center for the computing time made available
to perform the numerical tests. Special thanks to Daniel Wortmann and
the FLEUR team for providing the input files that generated the
eigenproblems used in the numerical tests.

\bibliography{numalg,comphys,mypab,soft}{}
\bibliographystyle{plain}

\end{document}

%% file: symacros.tex
\newcommand\eqdef{\buildrel {\rm def}\over=}
\newcommand\be{\begin{equation}}
\newcommand\ee{\end{equation}}
\newcommand\bea{\begin{eqnarray}}
\newcommand\eea{\end{eqnarray}}
\newcommand\bi{\begin{itemize}}
\newcommand\ei{\end{itemize}}

\newcommand\nr{{\it n}({\bf r})}
\newcommand\kv{{\bf k}}
\newcommand\rv{{\bf r}}

\newcommand\eqalign[1]{%
	\vcenter{%
		\normalbaselines \advance\baselineskip 5pt
		\advance\lineskip 5pt \tabskip=0pt
		\halign{%
			&\hfil $\displaystyle{##{}}$&
			$\displaystyle{{}##}$\hfil\cr
			#1\crcr
			}%
		}%
	}

\newcommand\dotline{\par\hbox to \hsize{\dotfill}\par}

\makeatletter
\def\befored@t#1.#2.#3;{#1}
\def\afterd@t#1.#2.#3;{#2}
\def\refhead#1{\edef\next{\ref{#1}}\expandafter\befored@t\next..;}
\def\reftail#1{\edef\next{\ref{#1}}\expandafter\afterd@t\next..;}
\def\lsim{\mathrel{\mathpalette\@versim<}}
\def\gsim{\mathrel{\mathpalette\@versim>}}
\def\@versim#1#2{\vcenter{\offinterlineskip
        \ialign{$\m@th#1\hfil##\hfil$\crcr#2\crcr\sim\crcr } }}
\newcommand\becomes[1]{\mathchoice{\becomes@\scriptstyle{#1}}
   {\becomes@\scriptstyle{#1}} {\becomes@\scriptscriptstyle{#1}}
   {\becomes@\scriptscriptstyle{#1}}}
\def\becomes@#1#2{\mathrel{\setbox0=\hbox{$\m@th #1{\,#2\,}$}%
        \mathop{\hbox to \wd0 {\rightarrowfill}}\limits_{#2}}}
\makeatother

%% file: body.tex
\section{Introduction} 
We present a methodological approach to solve for eigenpairs of
sequences of correlated dense eigenproblems arising in Density
Functional Theory (DFT). The novelty of this approach resides in the
use of approximate solutions in combination with a simple block
eigensolver based on polynomially accelerated subspace iteration. When
parallelized for distributed memory architectures this iterative
method is a viable alternative to conventional dense eigensolvers both
in terms of scalability and performance. Ultimately our approach will
enable the DFT specialists to simulate larger and more complex
physical systems.

Within the realm of condensed-matter physics, DFT is considered the
standard model to run accurate simulations of materials. The
importance of these simulations is two-fold: on the one hand they are
used to verify the correctness of the quantum mechanical
interpretation of existing materials. On the other hand, simulations
constitute an extraordinary tool to verify the validity of new
atomistic models which may ultimately lead to the invention of brand
new materials.

Each simulation consists of a series of self-consistent cycles; within
each cycle a fixed number $\mathcal{N}_{\kv}$ of independent
eigenvalue problems is solved. Since dozens of cycles are necessary to
complete one simulation, one ends up with $\mathcal{N}_{\kv}$
sequences made of dozens of eigenproblems. The properties of these
eigenproblems depend on the discretization strategy of the specific
DFT method of choice. In this paper we will exclusively consider the
Full-Potential Linearized Augmented Plane Waves method (FLAPW). This
DFT method gives rise to dense hermitian generalized eigenproblems
(DGEVP) with matrix size typically ranging from 2,000 to 20,000.

In FLAPW only a fraction of the lowest part of the eigenspectrum is
required. The eigenvalues inside this fraction correspond to the
energy levels below Fermi energy and their number never falls below
3\% or exceeds 20\% of the eigenspectrum. The relatively high number
of eigenpairs in combination with the dense nature and the size of the
eigenproblems inevitably lead to the choice of direct
eigensolvers. Direct eigensolvers follow a constrained path of linear
transformations starting from the generalized eigenproblem and
arriving to a tridiagonal one. In turn, the tridiagonal problem is
solved iteratively using one of the two methods available for
computing just a fraction of the spectrum, namely bisection inverse
iteration (BXINV)~\cite{Peters:1971vc} and multiple relatively robust
representations (MRRR)~\cite{Dhillon:1997to,Dhillon:2004hx}.

Until very recently, the computational strategy on parallel
distributed memory architecture favored the use of
ScaLAPACK~\cite{Blackford:1987ta} implementation of BXINV. Modern and
efficient dense libraries, like ELPA~\cite{Auckenthaler:2011cy} and
EleMRRR~\cite{Petschow:2012vc}, improve the performance but do not
change the overall computational strategy: each problem in the
sequence is solved in complete independence from the previous one. The
latter choice is based on the view that problems in the sequence are
apparently only loosely connected. In this paper we propose a
completely different strategy which tries to maximally exploit the
sequence of eigenproblems using an iterative eigensolver as opposed to
a direct one.

The novelty of our approach, in spite of the assumed loose connection
between eigenproblems, is in the use of the solutions of one problem
in the sequence as input when solving the next one. By its inherent
nature only an iterative method would be able to accept eigenvectors
as input. On the other hand not all such methods are capable of
maximally exploiting the information inputed. In this regards one of
the most effective methods is Subspace Iteration (SI). We have
implemented a version of this method accelerated with Chebyshev
polynomials. The end result is an algorithm (ChFSI) whose bulk of
computations is performed making use of the highly optimized Basic
Linear Algebra Subroutines (BLAS) library and can be easily
parallelized on shared and distributed memory architectures. In this
paper we present preliminary results for a distributed memory version
of ChFSI implemented using the Elemental library
framework~\cite{Poulson:2013fs}.

\section{FLAPW Simulations on Large Parallel Architectures}

Every DFT method is based on a variational principle stemming from the
fundamental work of Kohn and Hohenberg~\cite{Hohenberg:1964fz}, and its
practical realization~\cite{Kohn:1965zzb}. Central to DFT is the
solution of a large number of coupled one-particle Schr\"odinger-like
equations known as Kohn-Sham (KS).
\[
\left( \frac{\hbar^2}{2m}\nabla^2 
+ \mathcal{V}_{\rm eff}[\nr] \right) \phi_i(\rv) = E_i \phi_i(\rv)
\quad ; \quad \nr = \sum_i f_i \phi_i(\rv)
\]
Due to the dependence of the effective potential $\mathcal{V}_{\rm
  eff}$ on the charge density $\nr$, in itself a function of the
orbital wave functions $\phi_i(\rv)$, the KS equations are non-linear
and are generally solved self-consistently.

The KS equations need to be ``discretized'' in order to be
solved numerically. Intended in its broadest numerical sense, the
discretization translates the KS equations in a non-linear eigenvalue
problem. Eigenproblems generated by distinct discretization schemes
have numerical properties that are often substantially different; for
sake of simplicity we can group most of the schemes in three
classes. The first and the second classes make respectively use of
plane waves and localized functions to expand the one-particle orbital
wave functions $\phi_i(\rv)$ appearing in the KS equations
\begin{equation}
\label{eq:linearcomb}
\phi_i(\rv) \longrightarrow \phi_{\kv,i}(\rv) = 
\sum_{\bf G} c^{\bf G}_{\kv,i} \psi_{\bf G}(\kv,\rv).
\end{equation}
Methods in the third class do not use an explicit basis for the
$\phi_i(\rv)$'s but discretize the KS equations on a grid in real
space using finite differences. 

The eigenvalue problems emerging from the first two discretization
classes consist of dense matrices of small-to-moderate size while,
within real space methods, one ends up with very large sparse
matrices. Due to the dramatically different set of properties of the
eigenproblems, each DFT method uses a distinct strategy in solving for
the required eigenpairs. For instance it is quite common that methods
based on plane waves (ABINIT, VASP, PARATEC, Castep, \dots) use direct
eigensolvers while real space methods (PARSEC, GPAW, Octopus, \dots)
make use of iterative eigensolver based on Krylov- or Davidson-like
subspace construction. From the point of view of software packages for
distributed memory architectures, the choice between direct or
iterative eigensolvers leads respectively to the use of traditional
parallel libraries like ScaLAPACK or PARPACK~\cite{Lehoucq:1998uq}.

In this paper we deal with a specific instance of a plane wave method
which splits the basis functions support domain: in a spherical
symmetric area around each atom, $\psi_{\bf G}$ receive contributions
by augmented radial functions, while plane waves are supported in the
interstitial space between atoms. This discretization of the KS
equations -- known as FLAPW -- translates in a set of
$\mathcal{N}_{\kv}$ quite dense DGEVP
\[
\sum_{\bf G'} (A_\kv)_{\bf GG'}\ c^{\bf G'}_{\kv,i} = \lambda_{\kv,i} 
\sum_{\bf G'} (B_\kv)_{\bf GG'}\ c^{\bf G'}_{\kv,i},
\]
each one labeled by a value of the plane wave vector $\kv$. The role
of eigenvectors is played by the $n$-tuple of coefficients $c_{\kv,i}$
expressing the orbital wave functions $\phi_i$ in terms of the basis
wave functions $\psi_{\bf G}.$

The entries of each DGEVP matrix are initialized by evaluating
numerically a series of expensive multiple integrals involving the
$\psi_{\bf G}$s. Since we are dealing with non-linear eigenvalue
problems, each DGEVP has to be solved in a chain of self-consistent
cycles
\[
A_\kv\ c_{\kv,i} = \lambda_{\kv,i} 
B_\kv\ c_{\kv,i}\ \longrightarrow\ P^{(\ell)}_\kv :\ A^{(\ell)}_\kv\ c^{(\ell)}_{\kv,i} = \lambda^{(\ell)}_{\kv,i} 
B^{(\ell)}_\kv\ c^{(\ell)}_{\kv,i} \qquad (\ell = 1, \dots, N).
\]
All along the sequence the solutions of all $P^{(\ell-1)}_\kv$ are
used to initialize the new eigenproblems $P^{(\ell)}_\kv$. In
particular the eigenvectors $c^{(\ell-1)}_{\kv,i}$ are used to derive
the orbital functions $\phi_{\kv,i}^{(\ell-1)}$ which in turn
contribute to the charge density $n^{(\ell-1)}(\bf r)$. At the next
cycle $n^{(\ell-1)}(\bf r)$ contributes to modify the potential
$\mathcal{V}_{\rm ef{}f}$ which causes the functional form of the
$\psi^{(\ell)}_{\bf G}$s to change. 
These new basis function set directly determines the initialization of
the entries of $A^{(\ell)}_\kv$ and $B^{(\ell)}_\kv$ and indirectly
the new eigenvectors $c^{(\ell)}_{\kv,i}$. The result is a sequence
$\left\{P^{(1)}_\kv \dots P^{(N)}_\kv\right\}$ for each $\kv$ where
the eigenpairs $(\lambda^{(N)}_{\kv,i},c^{(N)}_{\kv,i})$ converged
within tolerance to the solution of the original non-linear
problem. In theory the chain of computations that goes from
$P^{(\ell-1)}_\kv$ to $P^{(\ell)}_\kv$ implies a connection between
eigenvectors of successive eigenproblems. In practice there is no
known  mathematical formalism that makes this connection
explicit. Correlation between the eigenvectors becomes evident only
numerically~\cite{DiNapoli:2012fk}.

When solving for an eigenvalue problem the first high level choice is
between direct and iterative eigensolvers. The first are in general
used to solve for a large portion of the eigenspectrum of dense
problems. The latter are instead the typical choice for sparse
eigenproblems or used to solve for just few eigenpairs of dense
ones. In FLAPW the hermitian matrices $A_\kv$ and $B_\kv$ are quite
dense, have size not exceeding 20,000, and each
$P^{(\ell)}_\kv$ is solved for a portion of the lower spectrum not
bigger than 20\%. Consequently, when each DGEVP is singled out from
the rest of the sequence, direct solvers are unquestionably the method
of choice. Currently, most of the codes based on FLAPW
methods~\cite{FLEUR,Wien2k, Exciting} use the algorithms BXINV or MRRR
directly out of the ScaLAPACK or ELPA library.


If the use of direct solvers is the obvious choice when each
$P^{(\ell)}_\kv$ is solved in isolation, the same conclusion may not
be drawn when we look at a the entire sequence of
$\left\{P^{(\ell)}_\kv\right\}$. In~\cite{DiNapoli:2012fk} it is shown how
the correlation between eigenvectors of successive DGEVPs becomes
manifest in the evolution of the angles $\theta^{(\ell)}_{\kv,i} =
\langle c^{(\ell-1)}_{\kv,i} , c^{(\ell)}_{\kv,i} \rangle$. In
particular the $\theta^{(\ell)}_{\kv,i}$ decrease almost monotonically
as a function of cycle index $\ell$, going from $\sim 10^{-1}$ down to
$\sim 10^{-8}$ towards the end of the sequence.

\begin{figure}[!htb]
\hspace*{-0.7cm}
\centering
  \subfigure[Distribution of computing time.]{
  \includegraphics[scale=0.4]{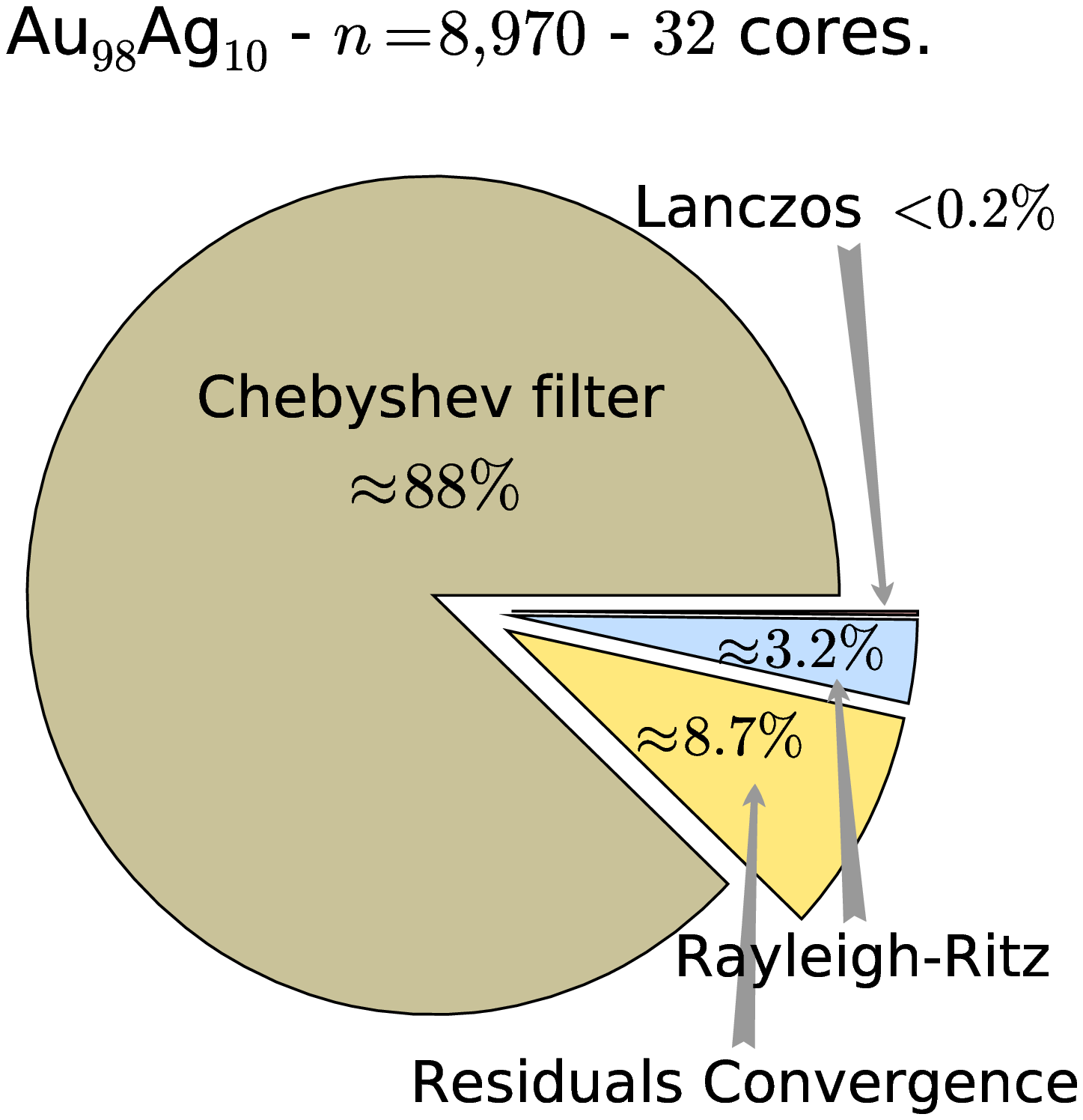}}
  \subfigure[Random vs Approximate vectors.]{
  \includegraphics[scale=0.4]{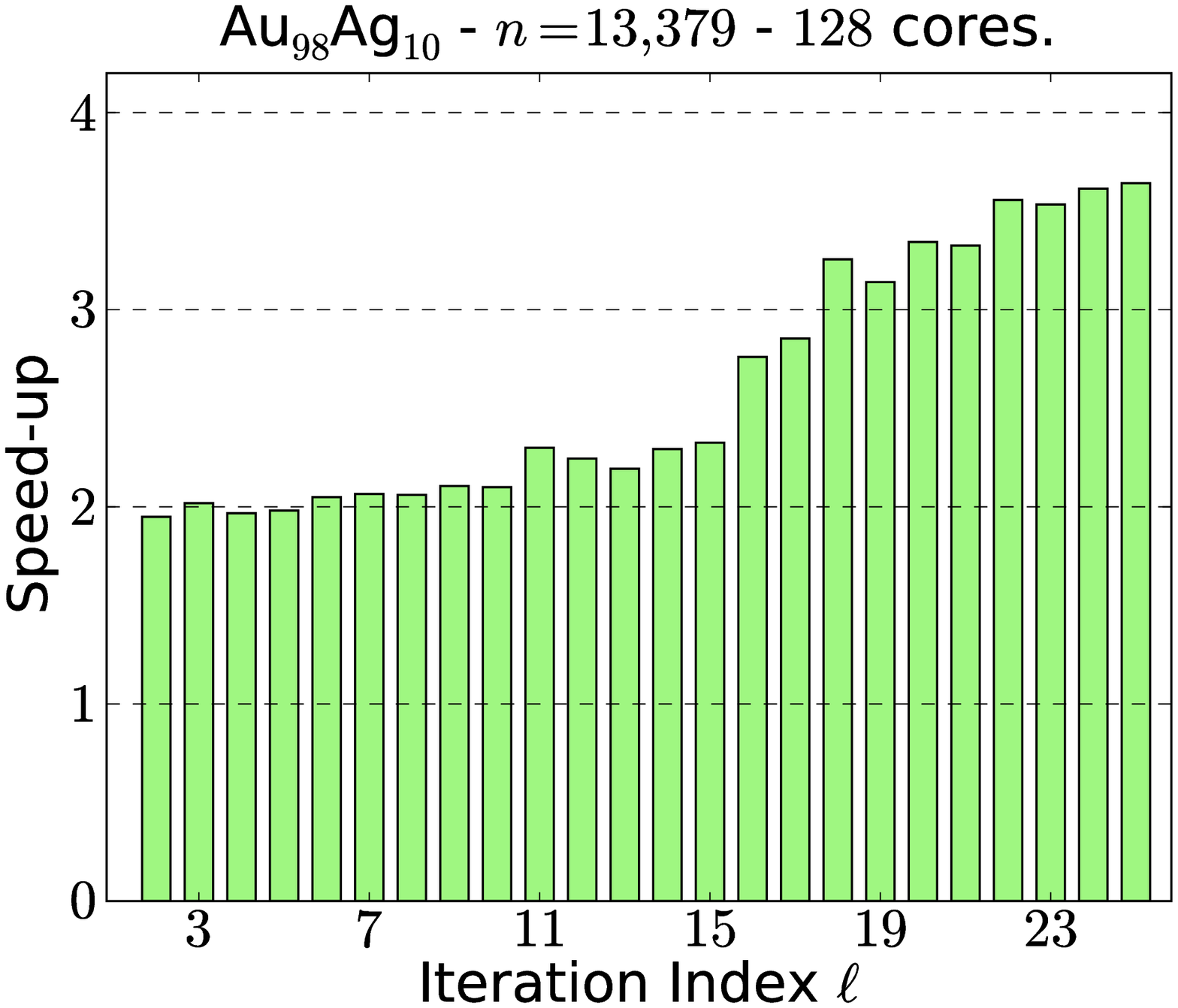}}
\caption{\it The data in this figure refers to eigenproblems of distinct
  sizes $n$ relative to the same physical system
  Au$_{98}$Ag$_{10}$. Plot (a) represents the computing fractions of
  EleChFSI's main algorithmic steps w.r.t. the total computing
  time. Plot (b) shows the speed-up of EleChFSI when inputed
  approximate solutions as opposed to random vectors.}
  \label{fig:piehist}
\end{figure}

The empirical evolution of the eigenvectors suggests that they can be
``reused'' as approximate solutions, and inputed to the eigensolver at
the successive cycle. Unfortunately no direct eigensolver is capable
of accepting vectors as approximate solutions. Therefore if we want to
exploit the progressive collinearity of vectors as the sequence
progresses, we are lead to consider iterative solvers; these solvers
by their own nature build approximate eigenspaces by manipulating
approximate eigenvectors. In particular we need a block iterative
eigensolver that accepts at the same time many vectors as input.
Among the many choices of block solvers, the Chebyshev Filtered
Subspace Iteration method (ChFSI) showed the highest potential to take
advantage of approximate eigenvectors~\cite{DiNapoli:2012uy}(see also
Fig.~\ref{fig:piehist}(b)). Since the core of the algorithm is based
on the repetitive use of matrix-matrix multiplications, the use of the
BLAS 3 library makes it very efficient and easy to scale.



\section{The Parallel Chebyshev Subspace Iteration}
Subspace Iteration complemented with a Chebyshev polynomial filter is a
well known algorithm in the literature~\cite{Saad:2011tu}. A version
of it was recently developed for a real space dicretization of DFT by
Chelikowsky {\sl et al.}~\cite{Zhou:2006ut,Zhou:2006ek} and included
in the PARSEC code~\cite{Kronik:2006ff}.

SI is probably one of the earliest iterative algorithms to be used as
numerical eigensolver. It is by definition a block solver since it
simply attempts to build an invariant eigenspace by multiplying a
block of vectors with the operator to be diagonalized.  It is a known
fact that any implementation based on subspace iteration converges
very slowly. By using a polynomial filter on the initial block of
inputed vectors the method experiences a high rate of
acceleration. Unfortunately the block of vectors spanning the
invariant subspace could easily become linearly dependent. In order to
avoid such an occurrence SI is usually complemented with some
re-orthogonalization procedure.

\begin{algorithm}[h!t]
  \caption{Chebyshev Filtered Subspace Iteration with locking}
  \label{alg:ChFSI}
  \begin{algorithmic}[1]
  \Require Matrix $H^{(\ell)}$ of the DGEVP reduced to standard form, approximate eigenvectors $\hat{Y}^{(\ell-1)} := \left[ \hat{y}^{(\ell-1)}_1, \ldots, \hat{y}^{(\ell-1)}_{\textsc{nev}}\right]$ and eigenvalues $\lambda^{(\ell-1)}_1$ and $\lambda^{(\ell-1)}_{\textsc{nev}+1}.$
  \Ensure Wanted eigenpairs $\left( \Lambda,Y\right).$
  \item[]
  \State Estimating the largest eigenvalue. \label{lst:line:lanczos} \Comment{{\sc Lanczos}}
  \Repeat
  \State Filtering the vectors $\hat{Y}=C_{m}(\hat{Y}).$ \label{lst:line:cheby} \Comment{{\sc Chebyshev filter}}
  \State Re-orthonormalizing $\hat{Y}.$ \label{lst:line:QR} 
  \Comment{{\sc QR algorithm}} 
  \State Computing Rayleigh quotient $G=\hat{Y}^{\dagger}H^{(\ell)}\hat{Y}.$ \label{lst:line:rrstarts} 
  \Comment{{\sc Rayleigh-Ritz} (Start)}
  \State Solving reduced problem $G\hat{w}=\lambda \hat{w}$ giving $\big(\hat{\Lambda}, \hat{W}\big).$ 
  \State Computing $\hat{Y} = \hat{Y}\hat{W}.$ \label{lst:line:rrends} \Comment{{\sc Rayleigh-Ritz} (End)}
  \For{$i=\textrm{converged} \to \textsc{nev}$ } \label{lst:line:dlstarts} \Comment{{\sc Deflation \& Locking} (Start)}
    \If{${\rm Res}(\hat{Y}(:,i),\hat{\Lambda}(i))<\textsc{tol}$}
    \State $\Lambda\gets \left[\Lambda\;\hat{\Lambda}(i)\right]$
    \State $Y\gets \left[Y\;\hat{Y}(:,i)\right]$ 
    \EndIf
    \EndFor \label{lst:line:dlends} \Comment{{\sc Deflation \& Locking} (End)}
  \Until{ converged $\geq$ \textsc{nev}}
\end{algorithmic}
\end{algorithm}

Our ChFSI algorithm is a slightly more sophisticated version of the
basic SI and is specifically tailored for DFT-like eigenproblems. The
whole algorithm is illustrated in the {\bf Algorithm~\ref{alg:ChFSI}}
scheme. Notice that the initial input is not the initial $P^{(\ell)}$
but its reduction to standard form $H^{(\ell)} =
L^{-1}A^{(\ell)}L^{-\textsc{T}}$ where $B^{(\ell)}=LL^{\textsc{T}}$,
and $\hat{Y}^{(\ell-1)}$ are the eigenvectors of $H^{(\ell-1)}$. ChFSI
uses few Lanczos iterations ({\tt line~\ref{lst:line:lanczos}}) so as
to estimate the upper limit of the eigenproblem
spectrum~\cite{Zhou:2011ic}. This estimate is necessary for the
correct usage of the filter based on Chebyshev
polynomials~\cite{Saad:2011tu}. After the Chebyshev filter step ({\tt
  line~\ref{lst:line:cheby}}) the resulting block of vectors is
re-orthonormalized using a simple QR algorithm ({\tt
  line~\ref{lst:line:QR}}) followed by a Rayleigh-Ritz procedure ({\tt
  line~\ref{lst:line:rrstarts}}). At the end of the Rayleigh-Ritz step
eigenvector residuals are computed, converged eigenpairs are deflated
and locked ({\tt line~\ref{lst:line:dlends}}) while the non-converged
vectors are sent again to the filter to repeat the whole procedure.

The Chebyshev polynomial filter is at the core of the algorithm. The
vectors $\hat{Y}$ are filtered exploiting the $3$-terms recurrence
relation which defines Chebyshev polynomials of the first kind
\begin{equation}
\label{eq:3term}
C_{m+1}(\hat{Y}) = 2\ H\ C_m(\hat{Y}) - C_{m-1}(\hat{Y}) \quad ; 
\quad C_m(\hat{Y}) \eqdef C_m(H)\cdot \hat{Y}. 
\end{equation}
This construction implies all operations internal to the filter are
executed through the use of ZGEMM, the most performant among BLAS 3
routines. Since roughly $90\%$ of the total CPU time is spent in the
filter (see pie chart in Fig.~\ref{fig:piehist}), the massive use of
ZGEMM makes ChFSI quite an efficient algorithm and potentially a very
scalable one.

The parallel MPI version of ChFSI (EleChFSI) is implemented within the
Elemental library, a framework for distributed memory dense linear
algebra. The core of the library is the two-dimensional cyclic
element-wise (``elemental'' or ``torus-wrap'') matrix distribution
(default distribution hereafter).  The $p$ MPI processes involved in
the computation are logically viewed as a two-dimensional $r\times c$
process grid with $p=r\times c.$ The matrix
$A=[a_{ij}]\in\mathbb{F}^{n\times m}$ is distributed over the grid in
such a way that the process $(s,t)$ owns the matrix
   \[
   A_{s,t}=
   \begin{pmatrix}
     a_{\gamma,\delta} & a_{\gamma,\delta+c} & \hdots\\
     a_{\gamma+r,\delta} & a_{\gamma+r,\delta+c} & \hdots\\
     \vdots & \vdots &
   \end{pmatrix},
   \]
   where $\gamma \equiv \left(s+\sigma_r\right)\mod r$ and $\delta
   \equiv \left(t+\sigma_c\right)\mod c$, and $\sigma_r$
   and $\sigma_c$ are arbitrarily chosen alignment parameters.
   
\begin{figure}[!htb]
\hspace*{-0.9cm}
\centering
  \includegraphics[scale=0.44]{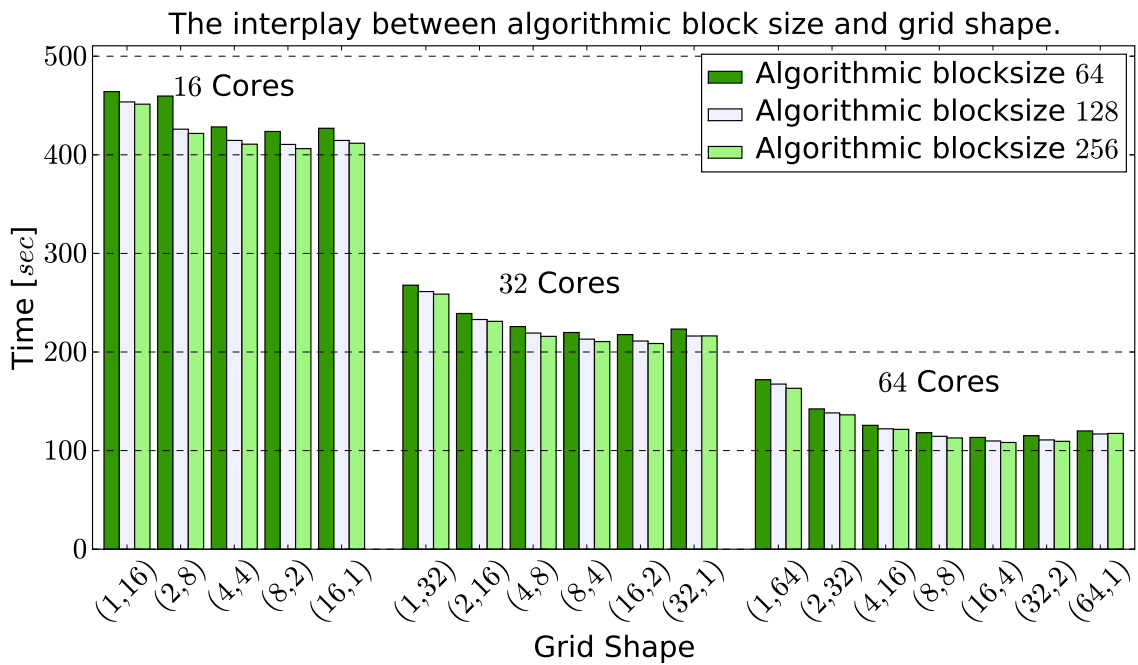}
  \caption{\it The data in this plot refer to a DGEVP of $\ell=20$, size
    $n=13,379$, and number of sought after eigenpairs $\textsc{nev} =
    972$, corresponding to the physical system Au$_{98}$Ag$_{10}$. The
    eigenproblem was repeatedly solved with EleChFSI using 16, 32, and
    64 cores, all possible grid shapes $(r,c)$ and three distinct
    algorithmic block sizes.}
  \label{fig:histogram}
\end{figure}

   For a given number $p>1$ of processors there are several possible
   choices for $r$ and $c$ forming dif{}ferent grid shapes
   $(r,c)\eqdef r\times c.$ Since the grid shape can have a
   significant impact on the overall performance, careful experiments
   should be undertaken in order to determine the best choice of
   $(r,c)$. Another parameter which affects performance is the
   algorithmic block size.  This term refers to the size of blocks of
   input data and is correlated to the square root of the L2
   cache~\cite{Goto:2008im}. In practice, the effective size of the
   algorithmic block not only depends on the algorithm itself, but it
   is also af{}fected by the architecture. Figure~\ref{fig:histogram}
   shows that for EleChFSI a block size of 256 is always recommended
   independently of the number of cores or grid shape. This effect is
   imputable to the large number of matrix multiplications carried on
   by the filter.
   
In the EleChFSI algorithm the Hamiltonian and the approximate
eigenvectors are distributed using the default distribution over the
$r\times c$ grid employing the Elemental library \texttt{DistMatrix}
class \footnote{The library provides several other matrix
  distributions \cite{Poulson:2013fs}.}  which internally ``hides''
the details about the matrix data-type, size, leading dimension, and
alignments. The net effect is to lift the user from the burden of
passing all those attributes to internal routines as it is customary
in (P)BLAS and (Sca/P)LAPACK libraries. The resulting separation of
concerns allows for the parallelization of the Chebyshev filter in a
straightforward fashion by calling the distributed memory
implementation of ZGEMM.  However, due to the generalization of the
3-term recursive relation, care must be taken with the distribution
update of diagonal entries of the Hamiltonian.

The reduced eigenproblem in the Rayleigh-Ritz step is solved using a
parallel implementation of the MRRR eigensolver --
EleMRRR~\cite{Petschow:2012vc} -- which is an integral part of
Elemental. The deflation and locking mechanism deserves particular
attention. When only a portion of the vectors are locked, the
algorithm has to re-filter a number of vectors that may, in general,
no longer have the same alignment $\sigma_c.$ To overcome this problem
the Elemental interface provides (among others) the routine
\texttt{View}, which takes as arguments two distributed matrices $A$
and $B$ and four integers $i, j, \mathit{height}$ and $\mathit{width}$
and makes $A$ a view of the $\mathit{height}\times\mathit{width}$
submatrix of $B$ starting at coordinate $(i, j).$ \footnote{The
  function is overloaded, and there are thus other dif{}ferent
  definitions.} The \texttt{View} routine works purely on pointers and
fully handles the distribution details eliminating the need of
allocating additional memory where to copy the data.

\begin{figure}[!htb]
\centering
  \subfigure[EleChFSI Strong scalability.]{
  \includegraphics[scale=0.39]{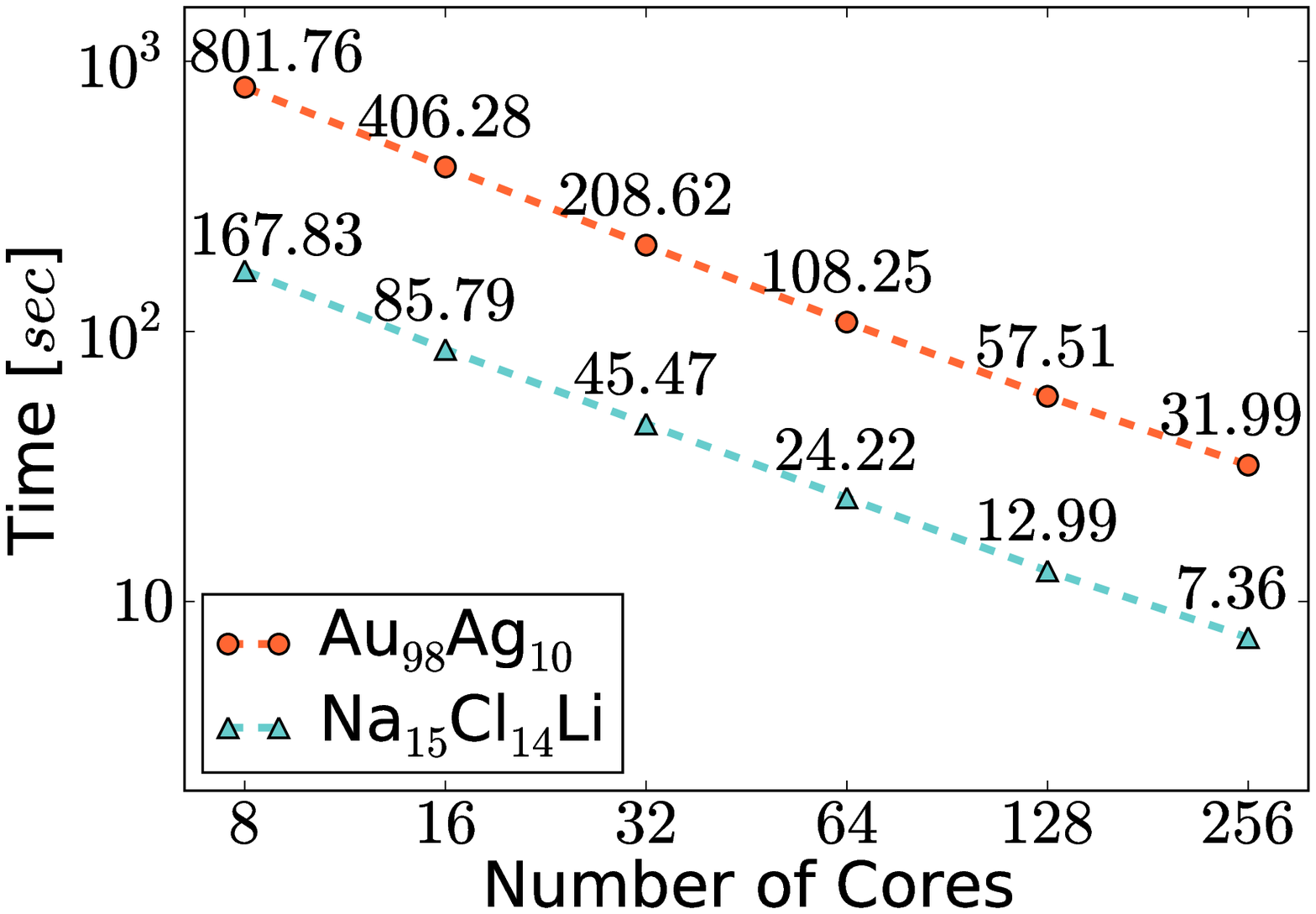}}
  \subfigure[EleChFSI Weak scalability.]{
  \includegraphics[scale=0.39]{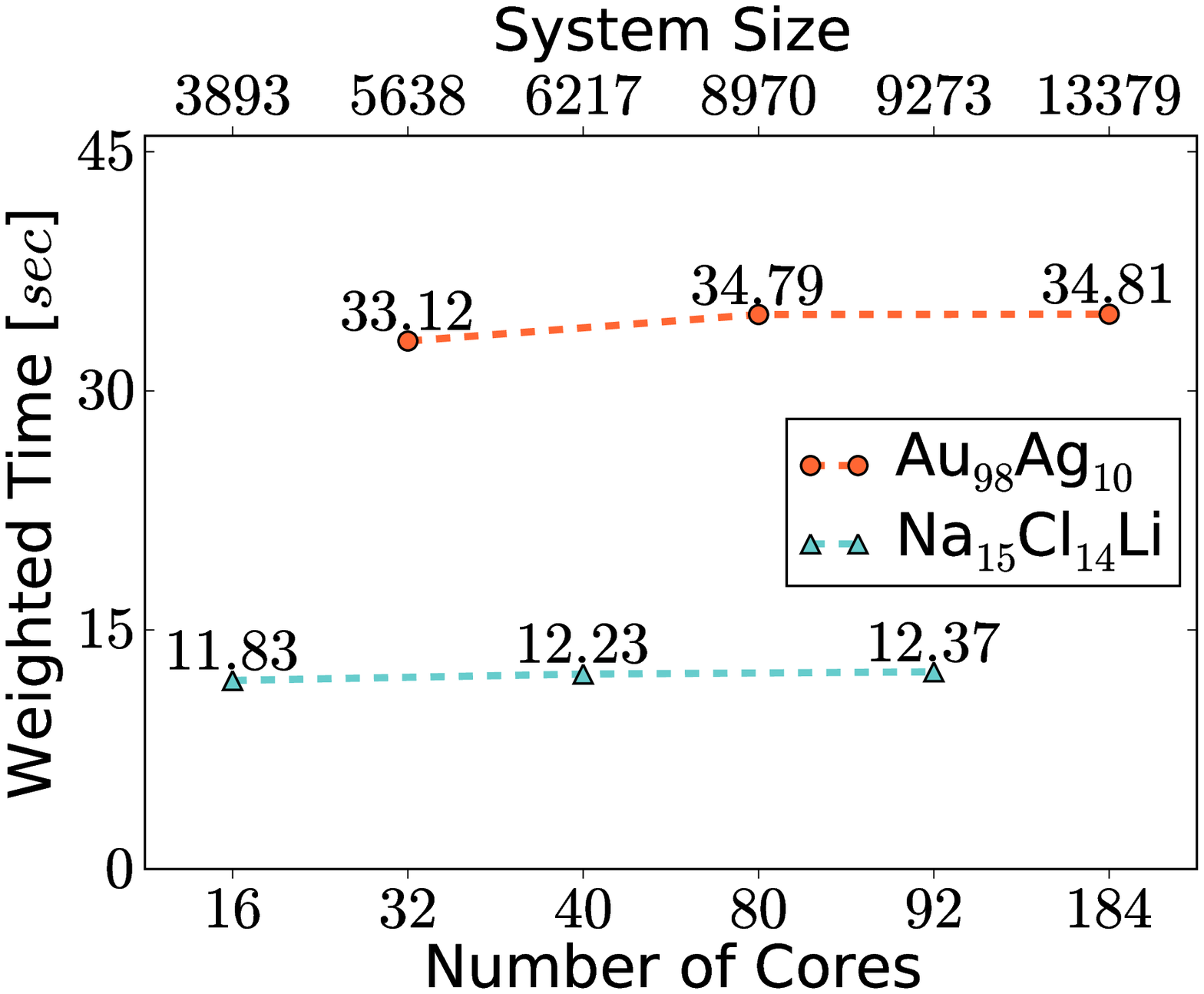}}
\caption{\it EleChFSI scalability for an increasing number of cores. In
  plot (a) the size of the eigenproblems are kept fixed while the
  number of cores is progressively increased. Eigenproblems of the two
  bigger system in Table \ref{tab:sim} are tested, namely $n=13,379$
  and $n=9,273$. In plot (b) all the systems are tested keeping the
  ratio of data per processor fixed. Times are weighted a posteriori
  by a factor keeping into account the ratio of operations per data
  varies in a non-predictable fashion with the size of the system.}
  \label{fig:scalability}
\end{figure}

The communication for the computations is performed almost entirely in
terms of collective communication within rows and columns of the
process grid. Such strategy in general implies that a square grid
shape is usually the best option~\cite{Chan:2007bm}. However, since in
our case we are solving for a small fraction of the eigenspectrum, the
matrix of vectors $\hat{Y}^{(\ell)}$ is tall and skinny. Consequently
we expect that a narrow rectangular grid shape will do a better job
than a square and wider one. This deduction is confirmed by
Fig.~\ref{fig:histogram}; independently of the number of cores the
optimal grid shape is either $(2^m,4)$ or $(2^{m+1},2)$, where $m >
2$.
   
\section{Numerical Results and Conclusions}

The set of numerical tests presented here were performed on two
distinct physical systems using three different sizes for the volume
of the reciprocal space defining the range of the vector {\bf G}
appearing in (\ref{eq:linearcomb}). Consequently we obtained three
sequences of eigenproblems for each physical systems. The data of the
sequences of eigenproblems are summarized in Table \ref{tab:sim}. All
our numerical tests were performed on JUROPA, a large general purpose
cluster where each node is equipped with 2 Intel Xeon X5570
(Nehalem-EP) quad-core processors at 2.93 GHz and 24 GB memory (DDR3,
1066 MHz).  The nodes are connected by an Infiniband QDR network with
a Fat-tree topology. The tested routines were compiled using the Intel
compilers (ver. 12.0.3) with the flag -O3 and linked to the ParTec's
ParaStation MPI library (ver. 5.0.26). The Elemental library was used
in conjunction with Intel's MKL BLAS (ver 11.0). All CPU times were
measured by running each test multiple times and taking the average of
the results. Eigenproblems were solved by EleChFSI by requiring the
eigenpairs residuals to be lower than $10^{-10}.$

\begin{table}[ht]
  \caption{ Simulation data}
  \centering
  \begin{tabular}{c c c c | c c c c }
  \hline \hline \\[-3mm]%
  Material & {\sc nev} & $\ell_{\rm max}$ & $n$ & Material & {\sc nev} & $\ell_{\rm max}$ & $n$\\ [0.5ex]
  \hline \\[-3mm]%
  \multirow{3}{*}{Au$_{98}$Ag$_{10}$} & \multirow{3}{*}{972} & 25 & 5,638 & \multirow{3}{*}{Na$_{15}$Cl$_{14}$Li} & \multirow{3}{*}{256} & 13 & 3,893\\ 
   & & 25 & 8,970 & & & 13 & 6,217\\
   & & 25 & 13,379 & & & 13 & 9,273\\ \hline%
  \end{tabular}\\
  \label{tab:sim}
  \end{table}
  As already mentioned in the previous sections,
  Fig.~\ref{fig:piehist} shows inequivocably the great advantage
  EleChFSI obtains from the use of the eigenvectors
  $\hat{Y}^{(\ell-1)}$ as input in solving the next eigenproblem
  $H^{(\ell)}$ in the sequence. This behavior is
  independent of the physical system or spectral properties of the
  eigenproblems: EleChFSI experiences speed-ups higher that 2X and
  often well above 3X towards the end of the sequence. Figure
  \ref{fig:histogram} also illustrate which is the optimal choice of
  grid shape and algorithmic block size. The remaining numerical tests
  were performed using exclusively the strategies outlined above.
\begin{figure}[!htb]
\centering
  \subfigure[Au$_{98}$Ag$_{10}$ - $n=13,379$]{
  \includegraphics[scale=0.39]{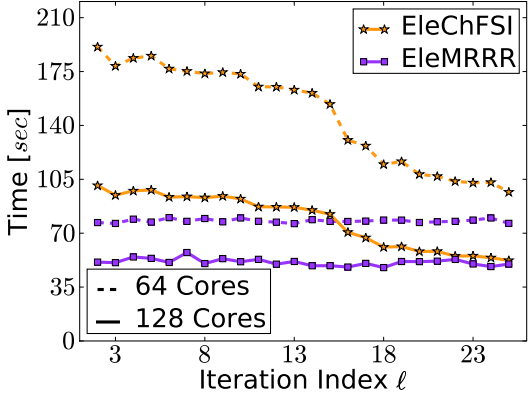}}
  \subfigure[Na$_{15}$Cl$_{14}$Li - $n=9,273$]{
  \includegraphics[scale=0.39]{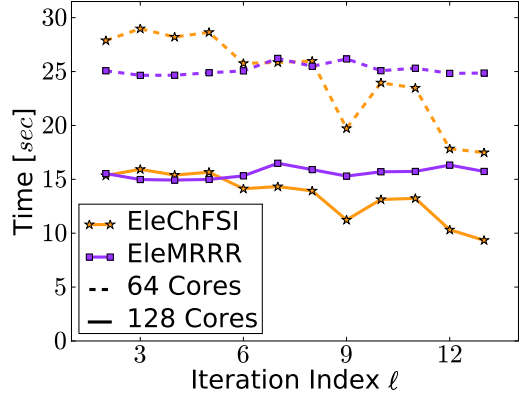}}
\caption{\it Comparing EleChFSI with EleMRRR on eigenproblems of
  increasing self-consistent cycle index $\ell$. For the size of
  eigenproblems here tested the ScaLAPACK implementation of BXINV is
  comparable with EleMRRR~\cite{Petschow:2012vc}. For this reason a
  direct comparison with the BXINV solver is not included here.}
\label{fig:direct}
\end{figure}

Fig.~\ref{fig:scalability} illustrate the scalability, both strong and
weak, of EleChFSI. Plot (a) shows a steady decrease of CPU time as the
number of cores increases. The rate of reduction is practically the
same for both systems despite their size differ by more than
$30\%$. This plot shows that EleChFSI is extremely efficient even when
the ratio of data per processor is not optimal. The weak scalability
plot makes manifest the great potential of this algorithm for
eigenproblems originating from FLAPW methods. The almost flatness of
the two lines implies that large size eigenproblems can greatly exploit
large supercomputing architectures. In other words EleChFSI has the
potential of allowing the users of FLAPW-based codes to generate more
complex physical systems made of thousands of atoms as opposed to just
few hundreds.

Compared to direct solvers, EleChFSI promises to be quite
competitive. Depending on the number of eigenpairs computed, our
algorithm is on par or even faster than EleMRRR. In plot (a) of
Fig.~\ref{fig:direct} EleChFSI appears to fall behind the direct
solver when using just 64 cores. The situation improves substantially
with 128 cores and at the end of the sequence both algorithms are on
par. The situation is even more favorable in plot (b) where EleChFSI
is already faster than EleMRRR for half of the eigenproblems in the
sequence (64 cores). When the tests are repeated with 128 cores
EleChFSI is inequivocably the faster of the two algorithms. Since the
fraction of the spectrum computed in plot (a) and (b) is respectively
$\sim 7\%$ and $\sim 3\%$, Fig.~\ref{fig:direct} shows that EleChFSI
scales better than EleMRRR and is more performant when the number of
eigenpairs is not too high.

In conclusion, not only EleChFSI showed to take the greatest advantage
from the progressive collinearity of eigenvectors along the sequence,
but it proved to easily adapt to parallel architectures. We showed how
such an algorithm, parallelized for distributed memory architectures,
scales extremely well
over a range of cores commensurate to the size of the
eigenproblems. Compared to direct eigensolvers, EleChFSI is
competitive with routines out of ScaLAPACK and Elemental. Eventually
the use of EleChFSI in FLAPW-based codes will enable the final user to
access larger physical systems which are currently out of reach.


%% file: main.bbl
\begin{thebibliography}{10}

\bibitem{Peters:1971vc}
Peters, G., Wilkinson, J.H.:
\newblock {The calculation of specified eigenvectors by inverse iteration}.
\newblock Handbook for Automatic Computation (1971)

\bibitem{Dhillon:1997to}
Dhillon, I.S.:
\newblock {A New O(n2) Algorithm for the Symmetric Tridiagonal
  Eigenvalue/Eigenvector Problem} (1997)

\bibitem{Dhillon:2004hx}
Dhillon, I.S., Parlett, B.N.:
\newblock {Multiple representations to compute orthogonal eigenvectors of
  symmetric tridiagonal matrices}.
\newblock Linear Algebra and its Applications \textbf{387} (August 2004)  1--28

\bibitem{Blackford:1987ta}
Blackford, L.S., Choi, J., Cleary, A., D'Azevedo, E., Demmel, J., Dhillon, I.,
  Dongarra, J., Hammarling, S., Henry, G., Petitet, A., Stanley, K., Walker,
  D., Whaley, R.C.:
\newblock {ScaLAPACK Users' Guide}.
\newblock Society for Industrial and Applied Mathematics (January 1987)

\bibitem{Auckenthaler:2011cy}
Auckenthaler, T., Blum, V., Bungartz, H.J., Huckle, T., Johanni, R.,
  Kr{\"a}mer, L., Lang, B., Lederer, H., Willems, P.R.:
\newblock {Parallel solution of partial symmetric eigenvalue problems from
  electronic structure calculations}.
\newblock Parallel Computing \textbf{37}(12) (December 2011)  783--794

\bibitem{Petschow:2012vc}
Petschow, M., Peise, E., Bientinesi, P.:
\newblock {High-Performance Solvers for Dense Hermitian Eigenproblems}.
\newblock arXiv preprint arXiv:1205.2107 (2012)

\bibitem{Hohenberg:1964fz}
Hohenberg, P.:
\newblock {Inhomogeneous Electron Gas}.
\newblock Physical Review \textbf{136}(3B) (November 1964)  B864--B871

\bibitem{Kohn:1965zzb}
Kohn, W., Sham, L.J.:
\newblock {Self-Consistent Equations Including Exchange and Correlation
  Effects}.
\newblock Phys.Rev. \textbf{140} (1965)  A1133--A1138

\bibitem{Lehoucq:1998uq}
Lehoucq, R.B., Sorensen, D.C., Yang, C.:
\newblock {ARPACK users' guide: solution of large-scale eigenvalue problems
  with implicitly restarted Arnoldi methods}.
\newblock (1998)

\bibitem{DiNapoli:2012fk}
Di~Napoli, E., Bl{\"u}gel, S., Bientinesi, P.:
\newblock {Correlations in sequences of generalized eigenproblems arising in
  Density Functional Theory}.
\newblock Computer Physics Communications \textbf{183}(8) (August 2012)
  1674--1682

\bibitem{FLEUR}
B\"ugel, S., Bihlmayer, G., Wortmann, D.:
\newblock {FLEUR}.
\newblock \url{http://www.flapw.de}

\bibitem{Wien2k}
Blaha, P., Schwarz, K., Madsen, G., Kvasnicka, D., Luitz, J.:
\newblock Wien2k.
\newblock \url{http://www.wien2k.at/}

\bibitem{Exciting}
:
\newblock {The Exiting Code}.
\newblock \url{http://exciting-code.org/}

\bibitem{DiNapoli:2012uy}
Di~Napoli, E., Berljafa, M.:
\newblock {Block Iterative Eigensolvers for Sequences of Correlated Eigenvalue
  Problems}.
\newblock preprint arXiv: 1206.3768v2 (June 2012)

\bibitem{Saad:2011tu}
Saad, Y.:
\newblock {Numerical methods for large eigenvalue problems}.
\newblock Siam (2011)

\bibitem{Zhou:2006ut}
Zhou, Y., Saad, Y., Tiago, M.L., Chelikowsky, J.R.:
\newblock {Parallel self-consistent-field calculations via Chebyshev-filtered
  subspace acceleration}.
\newblock Physical Review E \textbf{74}(6) (2006)  066704

\bibitem{Zhou:2006ek}
Zhou, Y., Saad, Y., Tiago, M.L., Chelikowsky, J.R.:
\newblock {Self-consistent-field calculations using Chebyshev-filtered subspace
  iteration}.
\newblock Journal of Computational Physics \textbf{219}(1) (November 2006)
  172--184

\bibitem{Kronik:2006ff}
Kronik, L., Makmal, A., Tiago, M.L., Alemany, M.M.G., Jain, M., Huang, X.,
  Saad, Y., Chelikowsky, J.R.:
\newblock {PARSEC -- the pseudopotential algorithm for real-space electronic
  structure calculations: recent advances and novel applications to
  nano-structures}.
\newblock physica status solidi (b) \textbf{243}(5) (April 2006)  1063--1079

\bibitem{Zhou:2011ic}
Zhou, Y., Li, R.C.:
\newblock {Bounding the spectrum of large Hermitian matrices}.
\newblock Linear Algebra and its Applications \textbf{435}(3) (August 2011)
  480--493

\bibitem{Poulson:2013fs}
Poulson, J., Marker, B., van~de Geijn, R.A., Hammond, J.R., Romero, N.A.:
\newblock {Elemental: A New Framework for Distributed Memory Dense Matrix
  Computations}.
\newblock ACM Transactions on Mathematical Software (TOMS) \textbf{39}(2)
  (February 2013)

\bibitem{Goto:2008im}
Goto, K., Geijn, R.A.v.d.:
\newblock {Anatomy of high-performance matrix multiplication}.
\newblock ACM Transactions on Mathematical Software \textbf{34}(3) (May 2008)
  1--25

\bibitem{Chan:2007bm}
Chan, E., Heimlich, M., Purkayastha, A., van~de Geijn, R.:
\newblock {Collective communication: theory, practice, and experience}.
\newblock Concurrency and Computation: Practice and Experience \textbf{19}(13)
  (2007)  1749--1783

\end{thebibliography}
